\documentclass[11pt,a4paper]{article}

\usepackage{jheppub}
\usepackage{graphicx}
\usepackage{dcolumn}
\usepackage{bm}
\usepackage{amsmath}
\usepackage{braket}
\usepackage{slashed}    
\usepackage{epstopdf}
\usepackage{placeins}
\usepackage{multirow}
\usepackage{makecell}
\usepackage[shortlabels]{enumitem}
\usepackage{cleveref}
\usepackage{xcolor}

\renewcommand\thesection{\arabic{section}}
\newcommand{\beq}{\begin{eqnarray}}
\newcommand{\eeq}{\end{eqnarray}}
\newcommand{\beqnn}{\begin{eqnarray*}}
\newcommand{\eeqnn}{\end{eqnarray*}}
\newcommand{\Tr}{\ensuremath{\mathrm{Tr}}}

\newcommand{\SU}{\mathrm{SU}}
\newcommand{\YM}{\mathrm{YM}}
\newcommand{\clov}{\mathrm{clov}}

\newcommand{\cool}{\mathrm{cool}}
\newcommand{\tor}{\mathrm{tor}}
\newcommand{\zpp}{0^{++}}
\newcommand{\G}{\mathrm{G}}

\begin{document}

\title{The $\theta$-dependence of the Yang--Mills spectrum from analytic continuation}

\author[a]{Claudio Bonanno,}
\emailAdd{claudio.bonanno@csic.es}
\affiliation[a]{Instituto de F\'isica T\'eorica UAM-CSIC, c/ Nicol\'as Cabrera 13-15, Universidad Aut\'onoma de Madrid, Cantoblanco, E-28049 Madrid, Spain}

\author[b]{Claudio Bonati,}
\emailAdd{claudio.bonati@unipi.it}
\affiliation[b]{Dipartimento di Fisica dell'Universit\`a di Pisa and INFN - Sezione di Pisa, Largo Bruno Pontecorvo 3, I-56127 Pisa, Italy}

\author[c]{Mario Papace,}
\emailAdd{mario.papace@uni-wuppertal.de}
\affiliation[c]{Fachbereich Mathematik und Naturwissenschaften, Bergische Universität Wuppertal, Gaußstraße 20, 42119 Wuppertal, Germany}

\author[d]{Davide Vadacchino}
\emailAdd{davide.vadacchino@plymouth.ac.uk}
\affiliation[d]{Centre for Mathematical Sciences, University of Plymouth, 2–5 Kirkby Place, Drake Circus, Plymouth, United Kingdom}

\abstract{We study the $\theta$-dependence of the string tension and of the
lightest glueball mass in four-dimensional $\mathrm{SU}(N)$ Yang--Mills
theories. More precisely, we focus on the coefficients parametrizing the
$\mathcal{O}(\theta^2)$ dependence of these quantities, which we investigate by
means of numerical simulations of the lattice-discretized theory, carried out
using imaginary values of the $\theta$ parameter. Topological freezing at large $N$ is avoided using the Parallel Tempering on Boundary Conditions algorithm. We provide controlled continuum extrapolations of such coefficients in the $N=3$ case, and we report
the results obtained on two fairly fine lattice spacings for $N=6$.}


\maketitle
\flushbottom

\section{Introduction}\label{sec:intro}

One of the most interesting features that emerges when studying the
non-perturbative regime
of quantum field theories (QFTs) is their
$\theta$-dependence: peculiar terms exist (the so called $\theta$-terms) which,
when added to the action, do not modify the classical equations of motion, 
and yet change the physical properties of the theory. The existence of
$\theta$-terms is related to the topological features of the gauge group and to
the space-time dimensionality~\cite{Coleman:1985rnk, Jackiw:1979ur}, so
$\theta$-dependence is not present in all QFTs. Nevertheless, several interesting
QFTs display a nontrivial $\theta$-dependence, ranging from Quantum
Chromo-Dynamics (QCD) and four-dimensional $\SU(N)$ Yang--Mills theories
~\cite{Coleman:1985rnk, Gross:1980br,Schafer:1996wv}, to two-dimensional models
like the $\mathrm{CP}^{N-1}$ models~\cite{DAdda:1978vbw,Witten:1978bc} and U($N$)
Yang--Mills theories~\cite{Kovacs:1995nn,Bonati:2019ylr,Bonati:2019olo} (and
even elementary quantum mechanical models~\cite{Jackiw:1979ur, Gaiotto:2017yup,
Bonati:2017woi}).

The vacuum energy (or the free energy, at finite temperature) is the physical
observable whose $\theta$-dependence has been investigated more thoroughly.  
The functional form of the vacuum energy in QCD can be estimated either
analytically in the chiral limit~\cite{DiVecchia:1980yfw, diCortona:2015ldu}
or perturbatively at the semi-classical level in the very 
high-temperature regime~\cite{Gross:1980br,Schafer:1996wv,Boccaletti:2020mxu}.
In the generic finite temperature case (or away from the chiral limit), the
coefficients of the Taylor's expansion of the free energy in powers of
$\theta^2$ can only be obtained through numerical simulations of the
lattice regularized theory~\cite{Bonati:2015vqz, Petreczky:2016vrs,
Frison:2016vuc, Borsanyi:2016ksw, Bonati:2018blm, Burger:2018fvb, Chen:2022fid,
Athenodorou:2022aay}. 
In the four-dimensional $\SU(N)$ Yang--Mills case,
lattice simulations are the main tool to study
the $\theta$-dependence
of the vacuum (or free) energy, and several lattice studies have been devoted
to investigating different aspects of this subject \cite{Alles:1996nm,
Alles:1997qe, DelDebbio:2004ns, DelDebbio:2002xa, DElia:2003zne, Lucini:2004yh,
Giusti:2007tu, Vicari:2008jw, Panagopoulos:2011rb, Bonati:2013tt, Ce:2015qha,
Ce:2016awn, Berkowitz:2015aua, Borsanyi:2015cka, Bonati:2015sqt,
Bonati:2016tvi, Bonati:2018rfg, Bonati:2019kmf}.
The large-$N$ limit is particularly interesting
as in this limit (at zero temperature), $\theta$-dependence is a
key ingredient in the Witten--Veneziano solution of the $\mathrm{U}(1)_{\mathrm{A}}$
problem~\cite{Witten:1979vv, Veneziano:1979ec,Kawarabayashi:1980dp}, and some general $N$-scaling
behaviors are theoretically expected~\cite{Witten:1980sp}. In two-dimensional
$\mathrm{CP}^{N-1}$ models, analytical predictions are available in the large-$N$ limit
for the coefficients of the Taylor expansion in $\theta^2$ of the vacuum
energy~\cite{DAdda:1978vbw, Campostrini:1991kv, DelDebbio:2006yuf,
Rossi:2016uce, Bonati:2016tvi}, which are nicely supported by numerical data~\cite{Vicari:1992jy, Bonanno:2018xtd, Berni:2019bch,
Bonati:2019olo, Bonati:2019ylr}. Finally, for two-dimensional $\mathrm{U}(N)$ Yang--Mills
theories we have complete analytic control of the $\theta$-dependence of the
vacuum energy~\cite{Kovacs:1995nn, Bonati:2019ylr, Bonati:2019olo}.

In this work we investigate an aspect of $\theta$-dependence that has
received far less attention: the $\theta$-dependence of the spectrum of the
theory. In QCD, close to the chiral limit, it is easy to derive the
$\theta$-dependence of the mass of the pseudo-Nambu--Goldstone bosons associated
to the spontaneous breaking of chiral symmetry~\cite{diCortona:2015ldu}; away from
the chiral limit, we have once again to resort to lattice simulations.
This is the case also for four-dimensional pure-gauge theories, which are
however much simpler to simulate than QCD. For this reason here we focus
on the case of four-dimensional $\SU(N)$ Yang--Mills theories, whose Euclidean
Lagrangian density is given by
\beq
\mathcal{L}_{\YM}(\theta) &=& \frac{1}{2g^2} \Tr\left\{G_{\mu\nu}(x)G_{\mu\nu}(x)\right\} + i \theta q(x)~,
\eeq
where
\beq
q(x) &=& \frac{1}{32 \pi^2} \varepsilon_{\mu\nu\rho\sigma} \Tr\left\{G_{\mu\nu}(x)G_{\rho\sigma}(x)\right\}~,
\eeq
and we investigate the $\theta$-dependence of the string tension $\sigma$ and
of the lightest glueball mass $m_{\G}$. Analytical computations can be performed
in two-dimensional $\mathrm{U}(N)$ Yang--Mills models, which however do not seem to provide 
much insight on the physics of their four-dimensional counterparts, since there is no 
$\theta$-dependence at all in the spectrum of these two-dimensional models in the continuum 
(see App.~\ref{sec:UN2d}).

At $\theta=0$, it is established
that the $\zpp$ glueball ground state represents the lightest state~\cite{Lucini:2004my, Athenodorou:2020ani, Athenodorou:2021qvs,
Vadacchino:2023vnc}.
At $\theta\neq 0$, spatial parity is explicitly broken, and cannot be used as 
a quantum number for glueball states.
The latter are thus only characterized by their spin and
charge conjugation quantum numbers. For this reason, we denote by $m_{\G}$ the mass
of the lightest glueball state, i.e., the one that 
tends to the $\zpp$ glueball in the $\theta\to0$ limit.
Note that, since in our study we only investigate 
the small-$\theta$ regime, we can a priori exclude the possibility of a
level crossing between different states. 

Using the invariance under parity of the
$\theta=0$ theory, we can parameterize the leading order $\theta$-dependence 
of the string tension and of the lightest glueball mass by the constants
$s_2$ and $m_2$, defined as follows:
\beq \label{eq:s2m2_def}
\sigma(\theta) &=& \sigma \left[1 + s_2 \theta^2 + \mathcal{O}(\theta^4)\right]\,, \\
m_{\G}(\theta) &=& m_{\zpp} \left[1 + m_2 \theta^2 + \mathcal{O}(\theta^4)\right]\,,
\eeq
where $\sigma$ and $m_{\zpp}$ stand for the string tension and the lightest
glueball mass computed at $\theta=0$, respectively. 
To the best of our knowledge, the only study in which an estimate of $m_2$
and $s_2$ was attempted is Ref.~\cite{DelDebbio:2006yuf}, where their values
have been obtained from the computation, at vanishing $\theta$, of the three-points 
correlation functions between the torelon or glueball interpolating operator 
and the square of the topological charge. As the calculation of 
these correlation functions is challenging, only results
of limited accuracy could be obtained in Ref.~\cite{DelDebbio:2006yuf}.

In this work we adopt an alternative approach, performing simulations at imaginary
values of $\theta$~\cite{Panagopoulos:2011rb}.
In a few words, assuming analyticity around $\theta=0$, we can
perform an analytic continuation from real to imaginary values of $\theta$,
which can be simulated without any sign problem. Since non analyticities
are expected to arise only for $\theta \sim \pi$, and since
we are dealing with the behavior of $\sigma$ and $m_{\G}$ at small $\theta$,
the analyticity assumption is well justified in our case and
poses no theoretical issues.
We can then 
obtain $\sigma(i\theta_I)$ and $m_{\G}(i\theta_I)$ 
directly for different values of
$\theta_I\in\mathbb{R}$, using the same standard algorithms used for
computations of the spectrum at $\theta=0$, and estimate
$s_2$ and $m_2$ from their small $\theta_I$ behavior.

Although this approach avoids the complications involved in the
evaluation of a three-point correlation function,
the computation of $s_2$ and $m_2$ remains
non-trivial, especially for large values of
$N$, 
where two main difficulties arise. The first is the rapid
increase of the integrated auto-correlation time of the topological 
modes in simulations
when approaching the continuum limit~\cite{Alles:1996vn,DelDebbio:2004xh, Schaefer:2010hu} (often referred to as the \emph{topological
freezing} problem), which becomes much stronger in the large-$N$ limit. To 
address this problem, we employ the Parallel Tempering on Boundary Conditions (PTBC)
algorithm~\cite{Hasenbusch:2017unr}, which has been shown to perform very well
in four-dimensional $\SU(N)$ gauge theories~\cite{Bonanno:2020hht}. The second
is that, as can be seen by using standard large-$N$
arguments~\cite{DelDebbio:2006yuf}, 
$s_2$ and $m_2$ are expected to scale as $1/N^2$.
Hence they are generically expected to have small values,
and to have a worsening of their
signal-to-noise ratio when the value of $N$ is increased.

As a final remark, we note that a reliable estimate of the coefficients $s_2$
and $m_2$ is not only important from a theoretical point of view, but is also
directly useful in numerical simulations.
As shown in~\cite{Brower:2003yx, Aoki:2007ka}, these coefficients describe the
systematical error that would be introduced by estimating the spectrum
from simulations at fixed topological charge $Q$. If $M$ is the
mass of a state and $M^{(Q)}$ is its estimate at fixed topological charge $Q$,
then
\beq \label{eq:Qbias}
\frac{M^{(Q)} -M}{M} \approx \frac{M_2}{2 \chi V}\ ,
\eeq
where $M_2$ is once again defined by $M(\theta)=M(1+M_2\theta^2+\cdots)$,
$\chi$ is the topological susceptibility and $V$ the space-time volume.
The coefficient $m_2$ can thus be used to impose
an upper bound on the
finite size effects introduced by a fixed topological background in the
computation of the $m_{\zpp}$ glueball mass.

This paper is organized as follows: in Sec.~\ref{sec:setup} 
we present our numerical setup, discussing the discretization adopted, the update algorithm and 
the procedure used to evaluate $\sigma$ and $m_\G$; in Sec.~\ref{sec:res} we present our numerical
results for the coefficients $m_2$ and $s_2$ parametrizing the $\theta$ dependence
of the string tension and of the lightest glueball state, discussing separately the cases $N=3$ and $N=6$; finally, in
Sec.~\ref{sec:conclu} we draw our conclusions and discuss some open problems. Two appendices report
the analytic computations performed in two dimensional $\mathrm{U}(N)$ models
and the tables with the raw numerical data of the four-dimensional $\SU(N)$ cases.

\section{Numerical setup}\label{sec:setup}

\subsection{Lattice discretization and simulation details}\label{sec:lat}

We discretize the $\SU(N)$ pure Yang--Mills theory at $\theta=0$ on an
isotropic hypercubic lattice with $L^4$ sites using the standard Wilson action:
\beq
S_{\mathrm{W}} = -\frac{\beta}{N} \sum_{x,\mu>\nu}\Re\,\Tr\left[\Pi_{\mu\nu}(x)\right],
\eeq
where $\Pi_{\mu\nu}(x)=U_{\mu}(x)U^\dagger_{\nu}(x+a\hat{\mu})U_{\mu}(x+a\hat{\mu})U_{\nu}(x)$
is the plaquette in position $x$ oriented along the directions $\mu\nu$, $a$ is the lattice spacing, and $\beta$ is the 
inverse lattice bare coupling. For the discretization of the topological charge we adopt 
the standard clover discretization
\beq
Q_{\clov} = \frac{1}{2^9 \pi^2} \sum_{\mu\nu\rho\sigma = \pm1}^{\pm4} 
\varepsilon_{\mu\nu\rho\sigma} \Tr\left[ \Pi_{\mu\nu}(x) \Pi_{\rho\sigma}(x) \right]\ ,
\eeq
in which $\varepsilon_{\mu\nu\rho\sigma}$ coincides with the
standard completely anti-symmetric tensor for positive values of the indices,
and its extension to negative values of the indices is uniquely fixed by
$\varepsilon_{(-\mu)\nu\rho\sigma}=-\varepsilon_{\mu\nu\rho\sigma}$ and
anti-symmetry. This definition ensures that $Q_{\clov}$ is odd under a lattice
parity transformation.  The action used to generate gauge configurations is
thus
\beq\label{eq:lattice_action_theta_imag}
S_L(\theta_L) = S_{\mathrm{W}} - \theta_L Q_{\clov},
\eeq
where the lattice parameter $\theta_L$ is related to the physical $\theta$
angle by $-i\theta = Z_Q \theta_L$, and $Z_Q$ is the (finite) renormalization
constant of the lattice topological charge~\cite{Campostrini:1988cy} $Q=Z_Q
Q_\clov$.

To estimate the numerical value of the renormalization constant $Z_Q$ it is
convenient to use smoothing algorithms such as cooling~\cite{Berg:1981nw, Iwasaki:1983bv,
Itoh:1984pr, Teper:1985rb, Ilgenfritz:1985dz, Campostrini:1989dh, Alles:2000sc}, 
smearing~\cite{APE:1987ehd, Morningstar:2003gk}, or gradient flow~\cite{Luscher:2009eq, 
Luscher:2010iy, Luscher:2011bx}, which dampen
the short-scale fluctuations while leaving the global topology of the
configurations unaltered. All smoothing algorithms have been shown to be
equivalent for this purpose~\cite{Alles:2000sc, Bonati:2014tqa,
Alexandrou:2015yba}, and in this work we adopt cooling to define an
integer-valued topological charge by using~\cite{DelDebbio:2002xa}
\beq
Q = \mathrm{round}\left\{\alpha \, Q_\clov^{(\cool)}\right\}\ ,
\eeq
with $\mathrm{round}\{x\}$ denoting the closest integer to $x$, and with $\alpha$ determined by the first nontrivial (i.e., $1<\alpha<2$) minimum of
\beq
\left\langle\left(\alpha Q_\clov^{(\cool)} - \mathrm{round}\left\{\alpha \, Q_\clov^{(\cool)}\right\}\right)^2\right\rangle \ .
\eeq
In this way we can determine the renormalization constant using~\cite{Panagopoulos:2011rb}
\beq\label{eq:renorm_Q}
Z_Q = \frac{\braket{Q Q_\clov}}{\braket{Q^2}}\ .
\eeq
Since the dependence of $Z_Q$ on the number $n_\cool$ of cooling steps used to
smooth the configurations is only very mild, reaching a plateau for
$n_\cool\sim 10-15$ for all the $\beta$ values studied, we define $Q$ using
$n_\cool=20$.

Simulations at imaginary values of the $\theta$ angle are by now recognized as
a cost-effective technique to study $\theta$-dependence on the lattice, as they
have been shown to typically outperform simulations carried out at
$\theta=0$~\cite{Bhanot:1984rx, Azcoiti:2002vk, Alles:2007br, Imachi:2006qq,
Aoki:2008gv, Panagopoulos:2011rb, Alles:2014tta, DElia:2012pvq, DElia:2013uaf,
Bonati:2015sqt, Bonati:2016tvi, Bonati:2018rfg, Bonati:2019kmf,
Bonanno:2018xtd, Berni:2019bch, Bonanno:2020hht, Bonanno:2023hhp}.  This is
especially true whenever $\theta=0$ simulations would require the computation
of higher-order (i.e., larger than two) correlators or susceptibilities, whose
order can be effectively reduced by performing simulations with an external
source, then studying the dependence of the results on the source strength. 
To determine the coefficients $s_2$ and $m_2$ at $\theta=0$ would require the
computation of three point functions, see Ref.~\cite{DelDebbio:2006yuf}, while
using simulations at $\theta\neq 0$ we can estimate $\sigma(\theta)$ and
$m_\G(\theta)$ as usual, from two-point functions. 
The values of $s_2$ and
$m_2$ are then determined from the behavior of $\sigma(\theta)$ and
$m_\G(\theta)$ for small $\theta$ values.  It should be clear that, in this
way, we are reducing the statistical errors.
However, we have to pay attention not to
introduce systematic ones, related to the determination of the
$\mathcal{O}(\theta^2)$ behavior of the observables for $\theta\approx 0$. More
details about the computation of $\sigma(\theta)$ and $m_\G(\theta)$ are
provided in the next subsection.

For simulations at $N=3$, we rely on the standard local updating algorithms
usually employed in pure-gauge simulations. More precisely, we adopt a 4:1
mixture of over-relaxation~\cite{Creutz:1987xi} and
heat-bath~\cite{Creutz:1980zw, Kennedy:1985nu} algorithms. For $N=6$, instead,
due to the severe topological freezing experienced by standard local algorithms
already at coarse lattice spacing, we adopt the Parallel Tempering on
Boundary Conditions (PTBC) algorithm, proposed for two-dimensional
$\mathrm{CP}^{N-1}$ models in Ref.~\cite{Hasenbusch:2017unr}. This algorithm
has indeed been shown to dramatically reduce the auto-correlation time of the
topological change both in two-dimensional models~\cite{Hasenbusch:2017unr,
Berni:2019bch, Bonanno:2022hmz} and in four-dimensional Yang--Mills
theories~\cite{Bonanno:2020hht,Bonanno:2022yjr,DasilvaGolan:2023cjw,Bonanno:2023hhp,Bonanno:2024nba}.

In a few words, in the PTBC algorithm $N_r$ replicas of the lattice theory in
Eq.~\eqref{eq:lattice_action_theta_imag} are simulated simultaneously. Each
replica differs from the others only by the boundary conditions imposed on a
small sub-region of the lattice, called \emph{the defect}; these boundary
conditions depend on a single parameter, which is used to interpolate between
periodic and open boundary conditions.  In this way a single replica has
periodic boundary conditions, another single replica has open boundary
conditions, and the intermediate $N_r-2$ replicas have ``mixed'' boundary
conditions, i.e., boundary conditions which interpolate between the two
previous ones. The state of each replica is updated using 
heat-bath and over-relaxation local updates, and configuration swaps 
between different replicas are proposed
during the MC evolution. These are accepted or rejected using a Metropolis
step. This algorithm allows to exploit the fast decorrelation of $Q$ achieved
with open boundaries~\cite{Luscher:2011kk}, avoiding at the same time the
difficulties related to the lack of translation invariance associated with the
presence of open boundaries. For more details on the implementation of this
algorithm we refer the reader to Ref.~\cite{Bonanno:2020hht}, where exactly the
same setup adopted here was used.

\subsection{Extraction of glueball and torelon masses}\label{sec:GEVP}

The starting point to evaluate the torelon mass, needed to extract 
the string tension, and the lightest glueball mass is the selection 
of a variational basis of zero-momentum-projected interpolating
operators $\mathcal{O}_i$. Each $\mathcal{O}_i$ is a (sum of) gauge invariant single-trace operators of fat-links, 
built by applying blocking and smearing algorithms to the lattice link
variables~\cite{BERG1983109, APE:1987ehd, TEPER1987345, Teper:1998te,
Lucini:2001ej, Lucini:2004my, Blossier:2009kd, Lucini:2010nv, Bennett:2020qtj,
Athenodorou:2020ani, Athenodorou:2021qvs}. For the computation of the lightest
glueball mass we employ 4-, 6- and 8-link operators in the $\mathrm{A}_1$
representation of the octahedral group, using a total of 160 operators. To
evaluate the torelon mass we use instead 5 operators, built in terms of 
products of fat-links winding around the time direction once. 

As noted before, the only discrete symmetry that can be used to classify the
states for non-vanishing values of $\theta$ is the charge conjugation C, since
parity is not conserved. For this reason, 
it would be natural to use a variational basis containing 
definite $\mathrm{C}$ only operators, without any projection
on definite $\mathrm{P}$ representations.
However, since 
we are just interested in
the properties 
of the ground state 
at small $\theta$ values, we can safely use the same
standard procedure adopted at $\theta=0$, i.e., use operators with definite parity. 
We have indeed verified that the extraction of the lowest glueball 
and torelon mass does not pose particular challenges 
and is always characterized by large enough overlaps ($A_\G>0.9$, with $A_\G$ the squared modulus of the matrix element between the ground state of the selected channel and the vacuum).

The optimal interpolating operator $\mathcal{O}=\sum_i v_i\mathcal{O}_i$ for
the ground state of the selected channel (i.e., the operator with the largest
value of $A_\G=\vert\langle 0 \vert \mathcal{O} \vert \G\rangle\vert^2$, with $\vert \G \rangle$ denoting the ground state) is the one whose weights
$v_i$ correspond to the components of the eigenvector of the Generalized
Eigenvalue Problem (GEVP)
\beq
C_{ij}(t) v_{j} = \lambda(t,t_0) C_{ij}(t_0) v_{j}, \qquad 
C_{ij}(t) \equiv \frac{1}{aL}\sum_{t^\prime}\braket{\mathcal{O}_i(t-t^\prime) \mathcal{O}_j(t^\prime)}
\eeq
associated to the largest eigenvalue $\overline{\lambda}$ (we typically used
$t_0/a=1$, performing also some checks using $t_0/a=2$). If we denote by $\bar{v}_i$ the
components of this eigenvector, the optimal correlator can be
written as
\beq
\overline{C}_{\G}(t) = C_{ij}(t) \overline{v}_i \overline{v}_j.
\eeq
The mass of the ground state is then obtained by fitting the functional form 
\beq\label{eq:fit_corr}
\overline{C}_{\G}(t) =  A_{\G}\left[ \exp\{-m_{\G} t\} + \exp\{-m_{\G}(aL-t)\}\right]
\eeq
in a range where the $t$-dependent effective mass
\beq\label{eq:meff_def}
a m_{\G}^{(\mathrm{eff})}(t) = - \log\left[\frac{\overline{C}_{\G}(t+a)}{\overline{C}_{\G}(t)}\right]
\eeq
exhibits a pleateau as a function of the time separation $t$. Final errors on
$a m_{\G}$ were estimated by means of a standard binned jack-knife analysis.

\section{Numerical results}\label{sec:res}

\subsection{Results for the \texorpdfstring{$\SU(3)$}{SU(3)} Yang--Mills theory}

For $N=3$ we performed simulations for 5 different values of the inverse
lattice bare coupling $\beta$, corresponding to lattice spacings ranging from
$\sim 0.1$~fm to $\sim 0.05$~fm. The lattice size $L$ was chosen large enough
to have $aL\sqrt{\sigma} \gtrsim 3.5$, in which case finite lattice size
effects are expected to be negligible to our level of precision, see, e.g.,
Ref.~\cite{Bonati:2015sqt}. As a further check that finite size effects are indeed
negligible, we compare our estimates of $m_{\zpp}/\sqrt{\sigma}$ at $\theta=0$
with the results of Ref.~\cite{Athenodorou:2020ani}, which have been obtained
using larger lattices (with $aL\sqrt{\sigma} \sim 4 - 5$), finding
perfect agreement.
 
\begin{figure}[!t]
\includegraphics[scale=0.435]{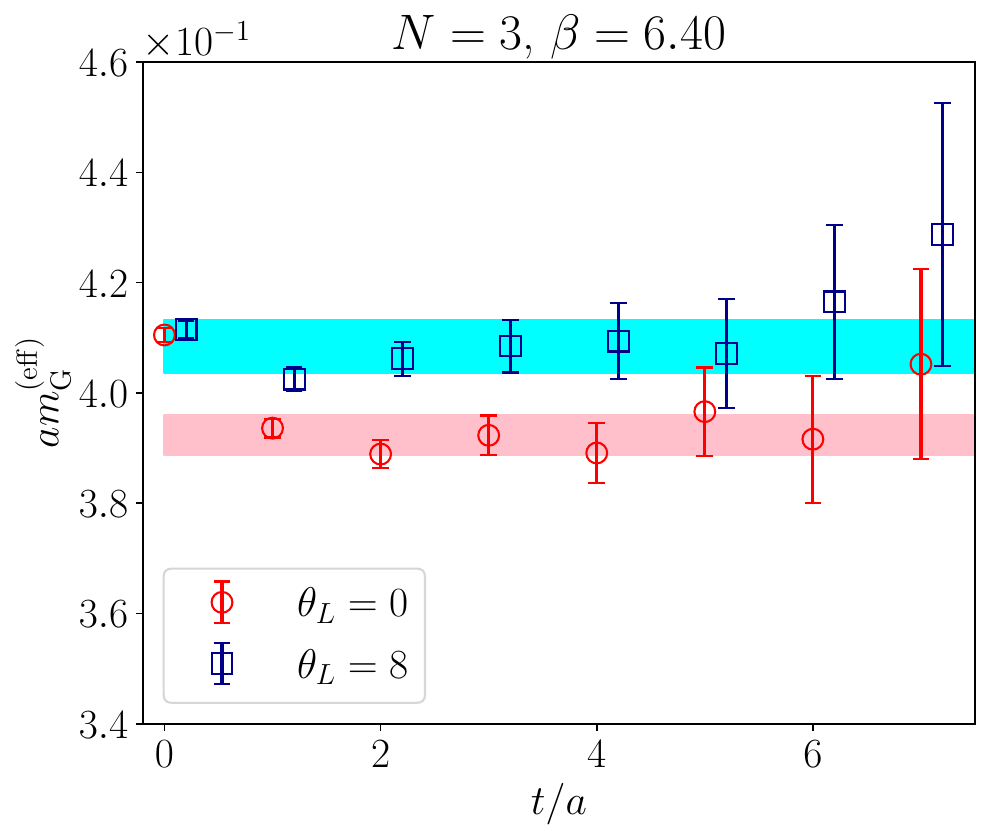}
\includegraphics[scale=0.435]{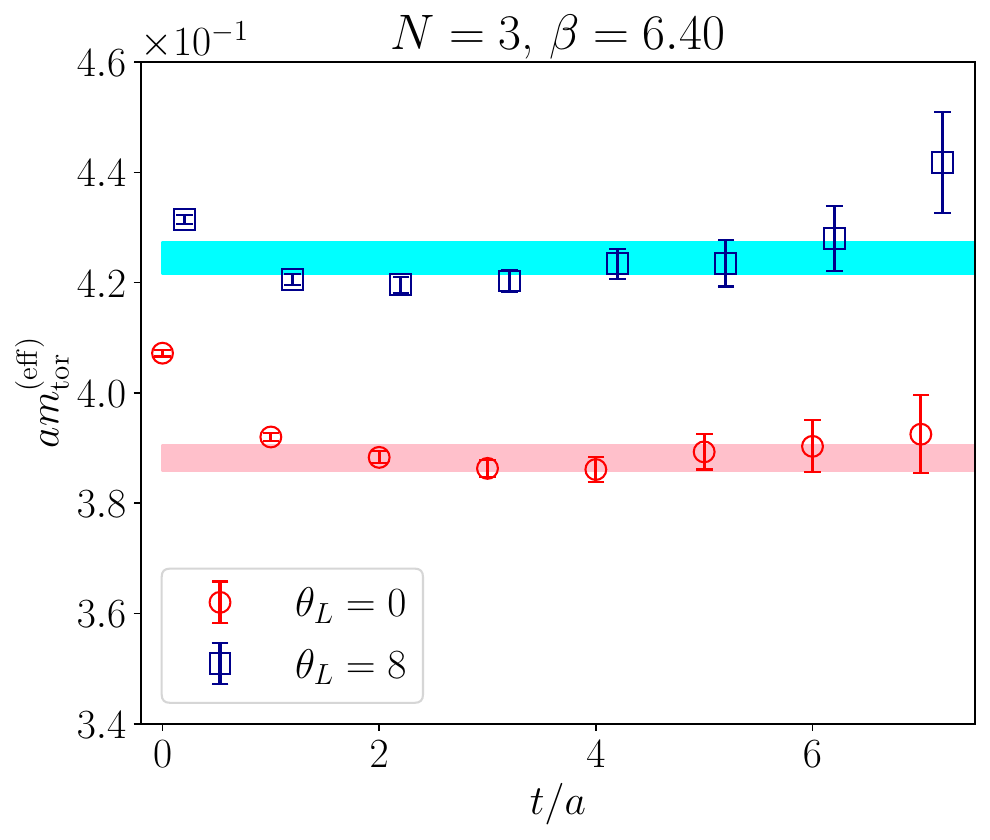}
\caption{Examples of $t$-dependent effective masses obtained for the lightest glueball (left) and torelon (right) states for $N=3$, $\beta=6.40$ and $\theta_L=0$ and $8$, computed using Eq.~\eqref{eq:meff_def} and the method described in Sec.~\ref{sec:GEVP}. Shaded bands represent our final results for $a_{\G}$ and $am_{\rm tor}$, obtained from a best fit of the optimal correlator according to Eq.~\eqref{eq:fit_corr} for $t/a\ge 3$, i.e., in the range where effective masses exhibit a plateau.}
\label{fig:meffs}
\end{figure}

Using the method described in Sec.~\ref{sec:GEVP}, we computed the lightest
glueball mass $m_{\G}$ and the torelon ground state mass $m_{\tor}$ for several
values of the lattice parameter $\theta_L$. For each value of $\beta$ and
$\theta_L$ we gathered a statistics of about $\mathcal{O}(60\text{k})$
thermalized configurations, separated from each other by 10 updating steps (1
step = 1 heat-bath and 4 over-relaxation sweeps of the whole lattice). The string
tension is extracted from the torelon ground state mass $m_\tor$ by the usual
formula~\cite{deForcrand:1984wzs}:
\beq\label{eq:tor_sigma_NLO}
a^2\sigma(\theta) = \frac{a m_{\tor}(\theta)}{L} + \frac{\pi}{3 L^2},
\eeq
and we explicitly verified that consistent results for $s_2$ (but not for
$\sigma(\theta)$) are obtained by using simply
$a^2\sigma(\theta)=am_{\tor}(\theta)/L$. For this reason, we used Eq.~\eqref{eq:tor_sigma_NLO}, including the next-to-leading correction, to compute $a^2\sigma$ for all values of $\theta_L$.  The estimates of $m_\G(\theta_L)$ and
$\sigma(\theta_L)$ thus obtained for all the probed values of $\beta$
and $\theta_L$ are reported for reference in
App.~\ref{sec:app_raw_data}. In Fig.~\ref{fig:meffs} we instead show a few examples of the obtained effective masses for the same coupling $\beta=6.40$ and for two values of $\theta_L=0$ and $8$.

The dependence of the lightest glueball mass and of the string tension on
$\theta_L$ can be parameterized, at leading order in $\theta_L$, as (see
Eq.~\eqref{eq:s2m2_def}):
\begin{equation}
\begin{aligned}
m_{\G}(\theta_L) &= m_{\zpp} \left[1 - \left(m_2 Z_Q^2\right) \theta_L^2 + \mathcal{O}(\theta_L^4)\right],\\
\sigma(\theta_L) &= \sigma \left[1 - \left(s_2 Z_Q^2\right) \theta_L^2 + \mathcal{O}(\theta_L^4)\right],
\end{aligned}
\end{equation}
where the relation $\theta^2 =- Z_Q^2\theta_L^2$ has been used, and $Z_Q$ is
the finite renormalization constant introduced in Sec.~\ref{sec:lat}. The
numerical value of $Z_Q$ depends on $\beta$, and we used the values reported in
Ref.~\cite{Bonati:2015sqt} in all but one case, namely $\beta=6.00$, in which
case $Z_Q$ has been estimated anew by using Eq.~\eqref{eq:renorm_Q} on
data at $\theta=0$ (as in Ref.~\cite{Bonati:2015sqt}).

\begin{figure}[!t]
\centering
\includegraphics[scale=0.435]{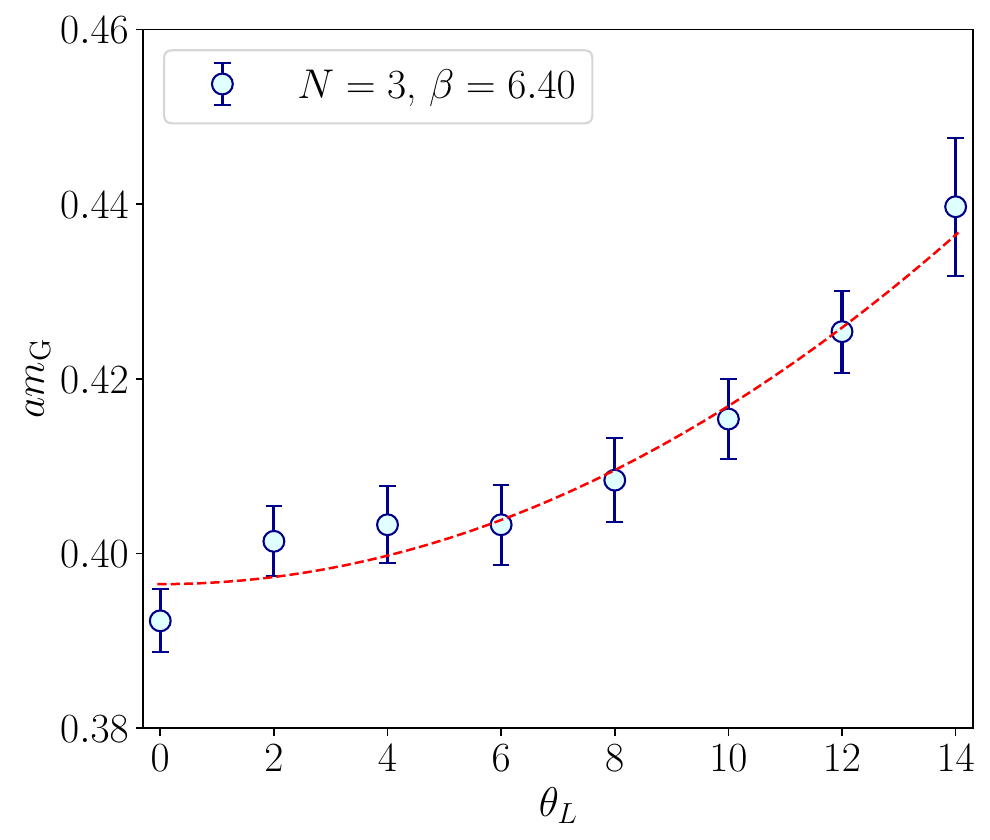}
\includegraphics[scale=0.435]{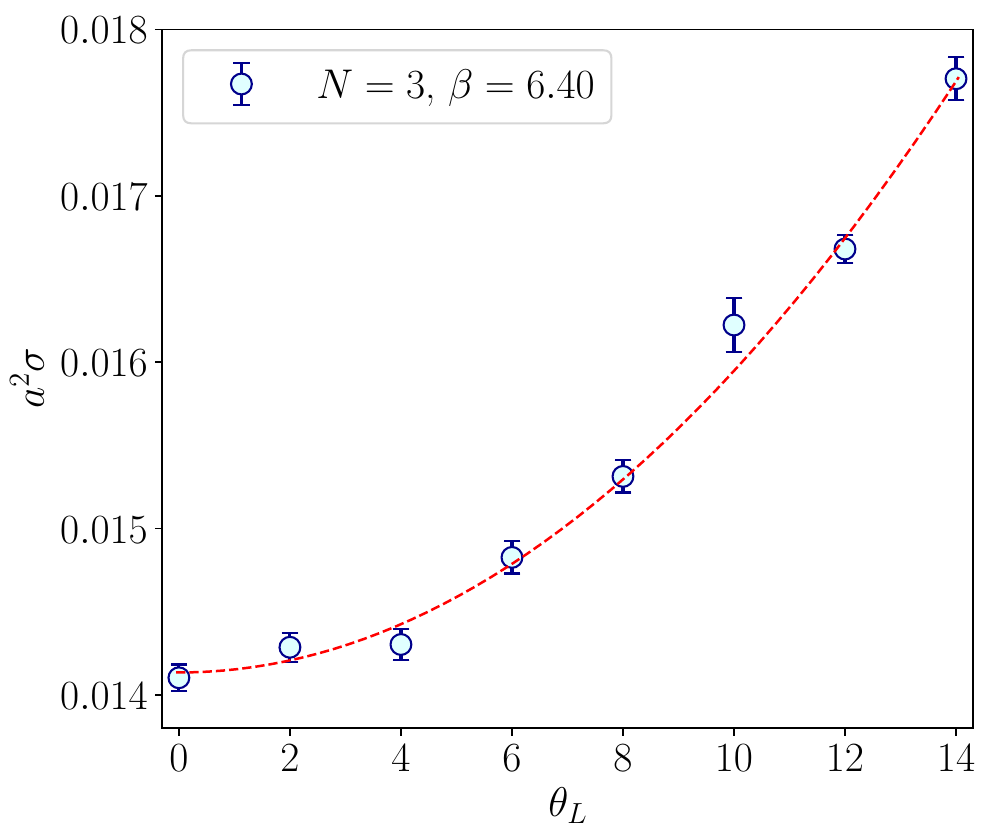}
\includegraphics[scale=0.435]{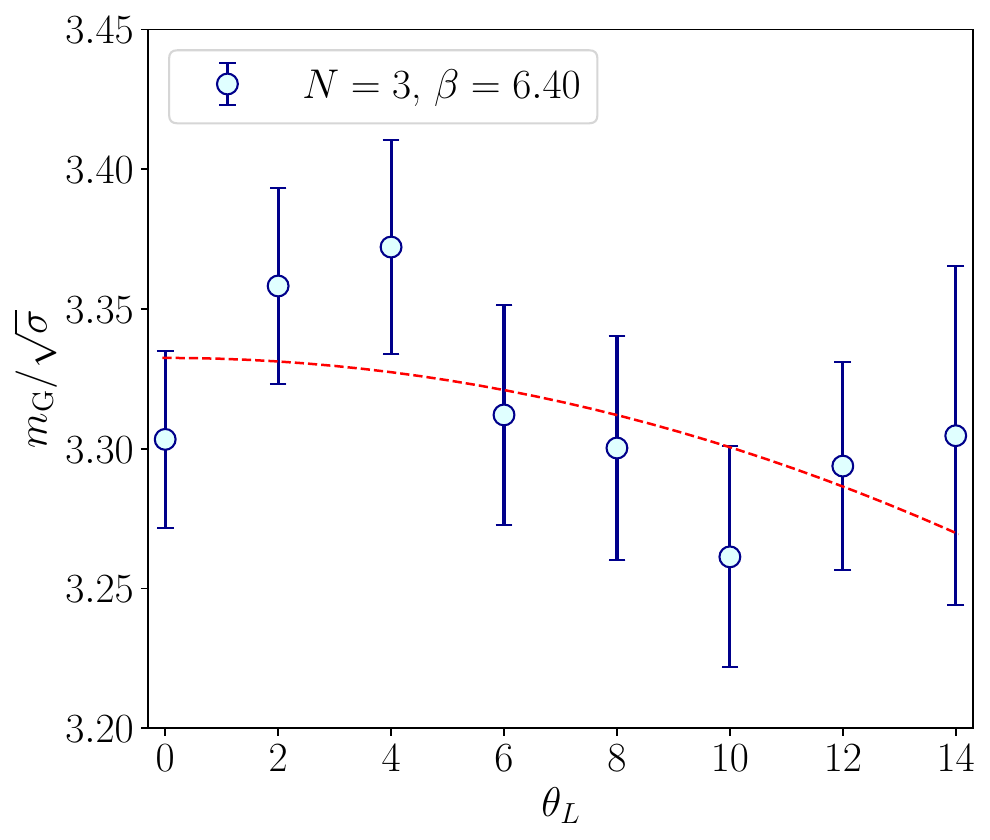}
\includegraphics[scale=0.435]{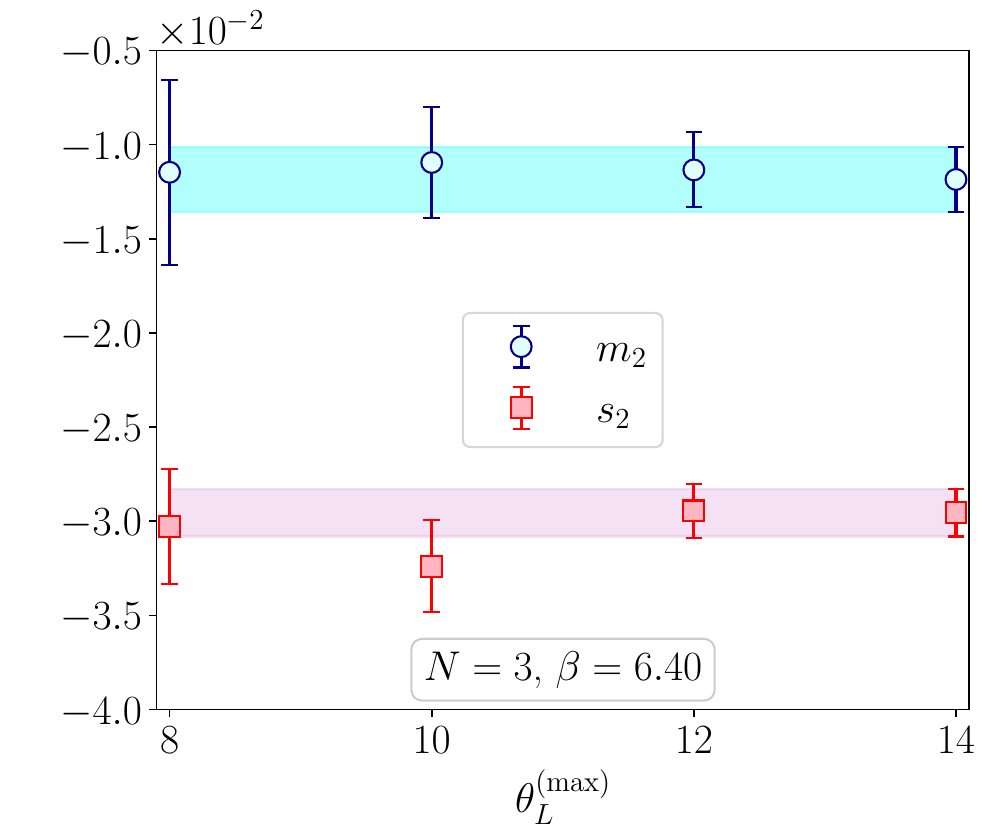}
\caption{Results for $N=3$ with $\beta=6.40$. Examples of fits performed by
using the functional form Eq.~\eqref{eq:fit_function} for several quantities:
(top left panel) the lightest glueball mass $a m_{\G}$, (top right panel) the
string tension $a^2 \sigma$, (bottom left panel) the ratio
$m_{\G}/\sqrt{\sigma}$. In the bottom
right panel we show the dependence of the estimates of $m_2$ and $s_2$ on the
upper limit of the fit range $\theta_L^{(\max)}$.}
\label{fig:ex_fit_theta_N3}
\end{figure}

To extract the values of $m_2$ and $s_2$ we performed a best fit of our data
for $am_{\G}(\theta_L)$ and $a^2\sigma(\theta_L)$ using the fit function
\beq\label{eq:fit_function}
f(\theta_L) = A_1[1+A_2 \theta_L^2 + \mathcal{O}(\theta_L^4)]\ ,
\eeq
where $A_1$ and $A_2$ are fit parameters. Examples of these fits are 
displayed in Fig.~\ref{fig:ex_fit_theta_N3}, 
from which it can be clearly seen that our data
are perfectly described by the leading $\mathcal{O}(\theta_L^2)$ behavior.  
To exclude the presence of systematical errors induced by the higher-order
$O(\theta_L^4)$ terms, we performed several fits, varying the upper limit
of the fit range.  When lowering the upper limit
of the fit range, the errors on the optimal fit parameters increase, but their central values remain well consistent with those obtained by using the full available range, as can be seen from the example shown in Fig.~\ref{fig:ex_fit_theta_N3}. For this reason we
report the results obtained by fitting all the available $\theta_L$ values.

\begin{table}[!t]
\begin{center}
\begin{tabular}{|c|c|l|c|c|l|c|}
\hline
$L$ & $\beta$ & \multicolumn{1}{c|}{$Z_Q$} & $a m_{\zpp}$ & $m_2$ & \multicolumn{1}{c|}{$a^2 \sigma$} & $s_2$ \\
\hline
16 & 5.95 &  0.12398(31)* &  0.7461(45) & -0.0247(28) & 0.05577(12) & -0.0426(11) \\
\hline
18 & 6.00 &  0.13554(39) &  0.6937(44) & -0.0190(31) & 0.04669(16) & -0.0419(20) \\
\hline
18 & 6.07 &  0.15062(62)* &  0.6220(35) & -0.0248(33) & 0.037079(89) & -0.0375(12) \\
\hline
22 & 6.20 &  0.1778(13)* &  0.5213(55) & -0.0172(34) & 0.02463(12) & -0.0363(20) \\
\hline
30 & 6.40 &  0.2083(29)* &  0.3965(22) & -0.0118(16) & 0.014135(46) & -0.0295(10) \\
\hline
\end{tabular}
\end{center}
\caption{Summary of the results obtained for $N=3$. The values of $Z_Q$ denoted
by an asterisk are from Ref.~\cite{Bonati:2015sqt}, while the value for
$\beta=6.00$ has been computed anew in this work.}
\label{tab:data_N3}
\end{table}

Our estimates of $m_2$ and $s_2$ for $N=3$ are summarized in
Tab.~\ref{tab:data_N3}. For comparison, in Ref.~\cite{DelDebbio:2006yuf} the
values of $s_2$ and $m_2$ were estimated by using simulations at $\theta=0$,
where $s_2=-0.077(15)$ and $m_2=-0.07(4)$ were obtained at $\beta=6.00$. 
Bearing in mind that the methods employed for these results are very different,
they appear to be in reasonable agreement. Moreover, they are
based on a roughly equivalent statistics, which shows that the improvement in 
accuracy is a benefit of the computational strategy used in the present work.

\begin{figure}[!t]
\centering
\includegraphics[scale=0.435]{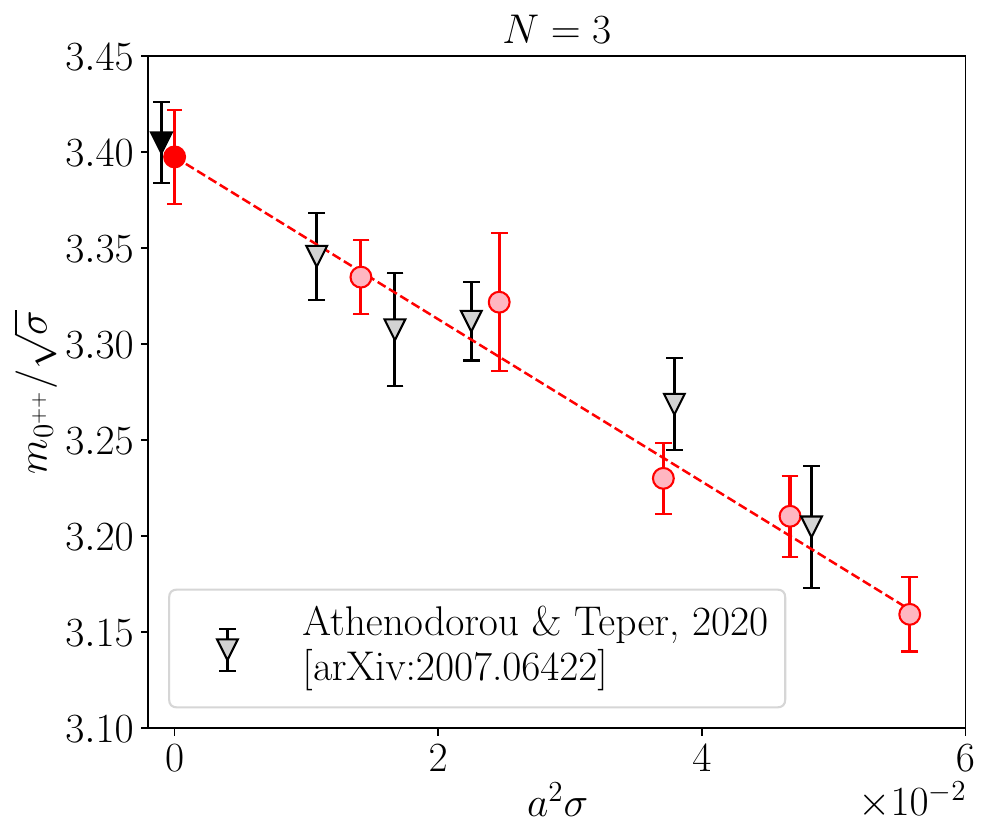}
\includegraphics[scale=0.435]{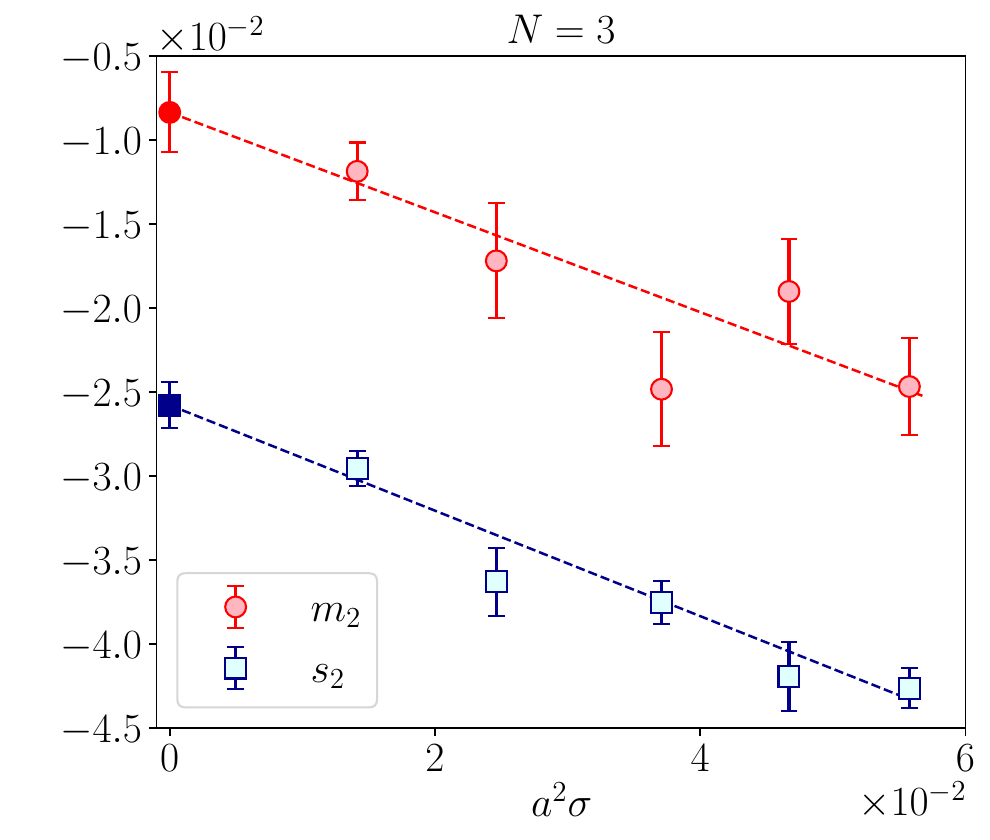}
\caption{Left panel: continuum scaling of our $N=3$ results for
$m_{\zpp}/\sqrt{\sigma}$, compared with data reported in
Ref.~\cite{Athenodorou:2020ani}. The dashed line is the result of a best fit of
our data assuming $\mathcal{O}(a^2)$ scaling corrections. Right
panel: continuum scaling of our $N=3$ results for $m_2$ and $s_2$ and their
continuum extrapolation assuming $\mathcal{O}(a^2)$ scaling corrections.}
\label{fig:ex_cont_limit_N3}
\end{figure}

Since $s_2<0$ the string tension increases when using simulations at
imaginary $\theta$, hence we do not expect to observe significant finite-size effects
at $\theta_L\neq 0$. As a further
check of the absence of finite-size effects we compared our results for the
ratio $m_{\zpp}/\sqrt{\sigma}$ (extracted from a fit of Eq.~\eqref{eq:fit_function}) 
with those obtained in Ref.~\cite{Athenodorou:2020ani} using significantly 
larger volumes. The comparison between these results is displayed in
Fig.~\ref{fig:ex_cont_limit_N3}, from which it is clear that they are
perfectly consistent with each other. In particular, assuming just $\mathcal{O}(a^2)$
corrections, we get the continuum limit
\beq
\frac{m_{\zpp}}{\sqrt{\sigma}} = 3.398(25)\ ,
\eeq
to be compared with $m_{\zpp}/\sqrt{\sigma} = 3.405(21)$ reported in Ref.~\cite{Athenodorou:2020ani}.

The continuum extrapolations of $m_2$ and $s_2$ are displayed in 
Fig.~\ref{fig:ex_cont_limit_N3}. These results are
obviously consistent with the
presence of just the leading $\mathcal{O}(a^2)$
finite-$a$ corrections. 
We thus obtain the continuum extrapolated values
\beq
\label{eq:m2N3}
m_2= -0.0083(23), \qquad \mathrm{(continuum\ extrapolated)},\\
\label{eq:s2N3}
s_2= -0.0258(14), \qquad \mathrm{(continuum\ extrapolated)}.
\eeq
Remarkably, the continuum extrapolated value of $s_2$ is quite close to twice
$m_2$, which means that the dimensionless ratio $m_\G(\theta)/\sqrt{\sigma(\theta)}$ is almost
independent of $\theta$. If we define $g_2$ by the equation
\begin{equation}
\frac{m_\G(\theta)}{\sqrt{\sigma(\theta)}}=
\frac{m_{\zpp}}{\sqrt{\sigma}}[1+g_2\theta^2+\mathcal{O}(\theta^4)],
\end{equation}
we indeed have the continuum result\footnote{We assume the statistical errors on $s_2$
and $m_2$ to be statistically independent, which is a reasonable guess since
they come from very different channels.}:
\beq
g_2 = m_2 - \frac{s_2}{2} = 0.0046(24), \qquad \mathrm{(continuum\ extrapolated)}.
\eeq
That the ratio $m_\G(\theta)/\sqrt{\sigma(\theta)}$ is quite insensitive to the
value of $\theta$ is also true at finite lattice spacing, as can be appreciated
from the data reported in Tab.~\ref{tab:data_N3} and from the example displayed in
Fig.~\ref{fig:ex_fit_theta_N3}.

Although we are not aware of any physical argument implying the vanishing of 
$g_2$, this result could suggest that all dimensionless quantities 
are
independent of $\theta$.  Such a strong statement can be however shown to be false.
In Ref.~\cite{DElia:2012pvq} (see also \cite{DElia:2013uaf, Bonanno:2023hhp}), 
the $\theta$-dependence of the $\SU(3)$ deconfinement critical temperature
$T_c$ was studied, and it was concluded that
\beq
T_c(\theta) = T_c[1 - R \theta^2 + \mathcal{O}(\theta^4)],
\eeq
where $T_c$ is the $\theta=0$ critical temperature and
\beq
R = 0.0178(5).
\eeq
This can be recast in units of $\sqrt{\sigma(\theta)}$ as:
\beq
\frac{T_c(\theta)}{\sqrt{\sigma(\theta)}} = 
\frac{T_c}{\sqrt{\sigma}}[1 -t_2\theta^2+ \mathcal{O}(\theta^4)],
\eeq
with (using Eq.~\eqref{eq:s2N3})
\beq
t_2 = R+\frac{s_2}{2} = 0.0049(9),
\eeq
which is definitely different from zero.

\subsection{Results for the \texorpdfstring{$\SU(6)$}{SU(6)} Yang--Mills theory}

The general strategy adopted at $N=6$ is the same
as for the case $N=3$. 
However, obtaining precise results in this case has proven to be a much
more challenging task.
We thus focused on just two
values of the bare inverse lattice coupling, namely $\beta=25.056$ and
$\beta=25.452$, corresponding to quite fine lattice spacings. 
Obviously, using estimates at only two values of the lattice spacing
prevents us from performing a reliable continuum extrapolation in this case.

The use of the PTBC algorithm was instrumental in reducing the auto-correlation
time of Monte Carlo simulations. In particular, for the two values of $\beta$
considered in this work, the PTBC algorithm allows to reduce
the integrated auto-correlation time of the topological modes
 (at constant CPU time) by a factor of $\sim 20$ for
$\beta=25.056$~\cite{Bonanno:2020hht}, and by a factor of $\sim 60$ for
$\beta=25.452$~\cite{Bonanno:2022yjr}. In simulations performed at inverse
coupling $\beta=25.056$ we produced and stored $\mathcal{O}(20\text{k})$ 
thermalized
configurations at $\theta=0$ and $\mathcal{O}(13\text{k})$ configuration for
each non-zero $\theta$ value, while for simulations at inverse coupling
$\beta=25.452$ we produced and stored $\mathcal{O}(5\text{k})$ 
thermalized configurations for
each non-zero value of $\theta$ (for $\theta=0$ we used results from
a previous study, see Ref.~\cite{Bonanno:2022yjr}).  
In all cases, measurements were performed every $10$
parallel tempering steps, using the same setup already adopted in
Refs.~\cite{Bonanno:2020hht,Bonanno:2022yjr}, to which we refer for more
details.

\begin{figure}[!t]
\centering
\includegraphics[scale=0.435]{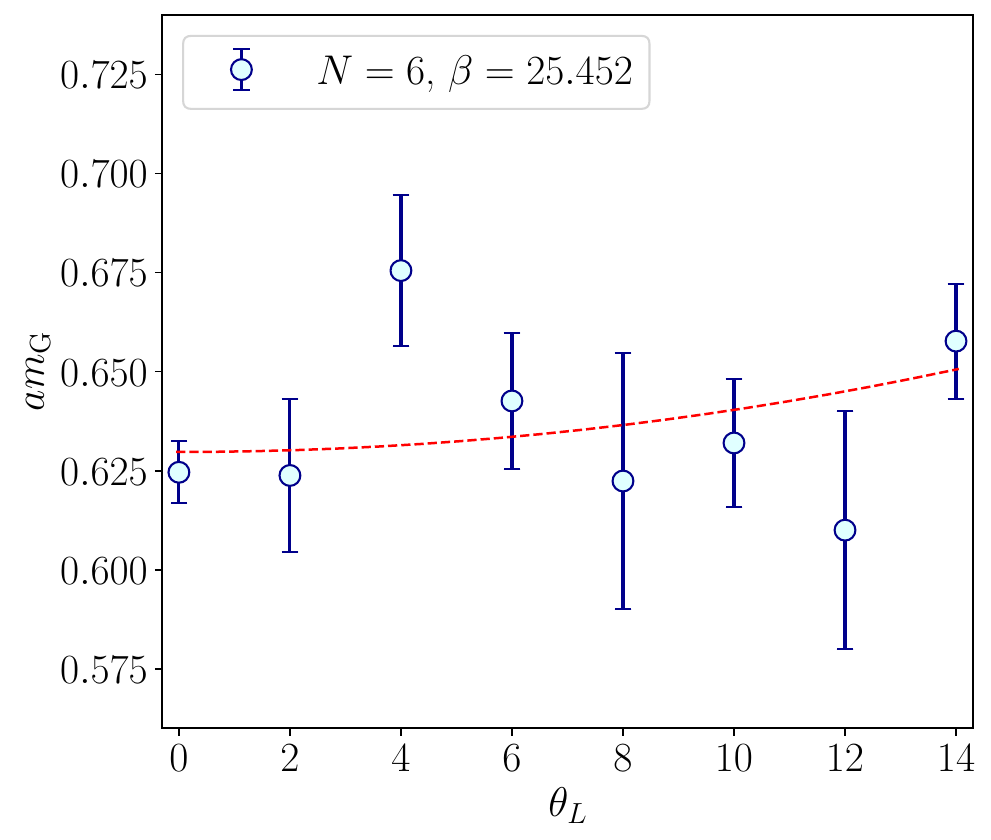}
\includegraphics[scale=0.435]{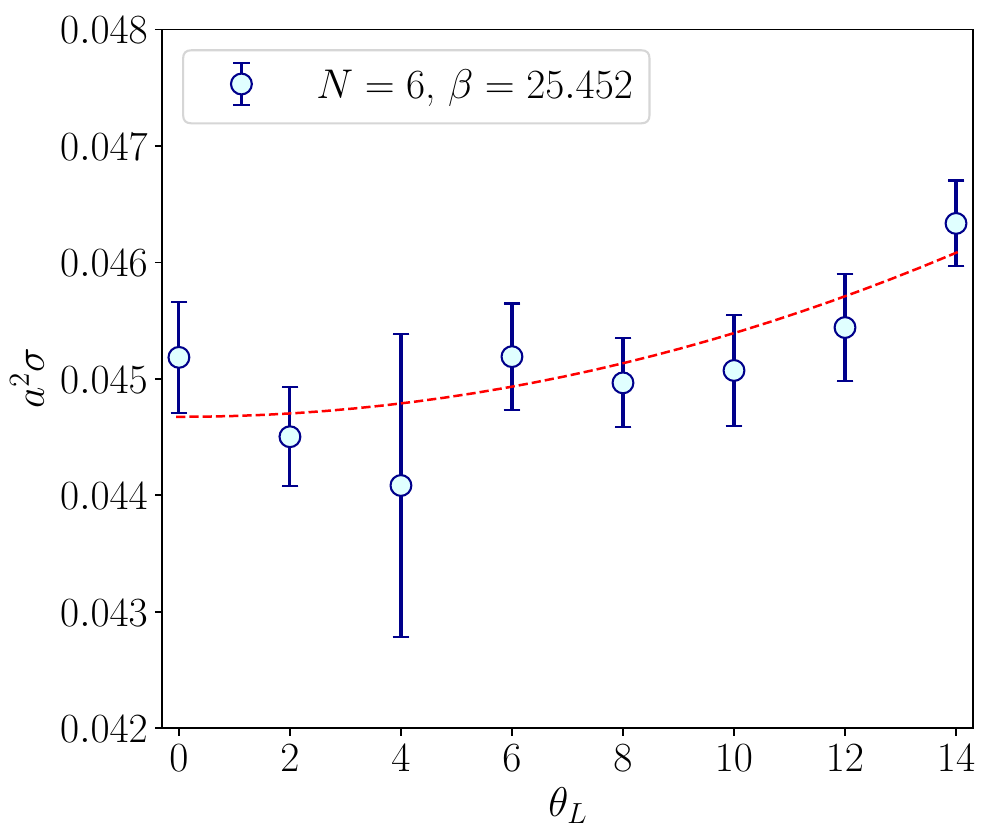}
\includegraphics[scale=0.435]{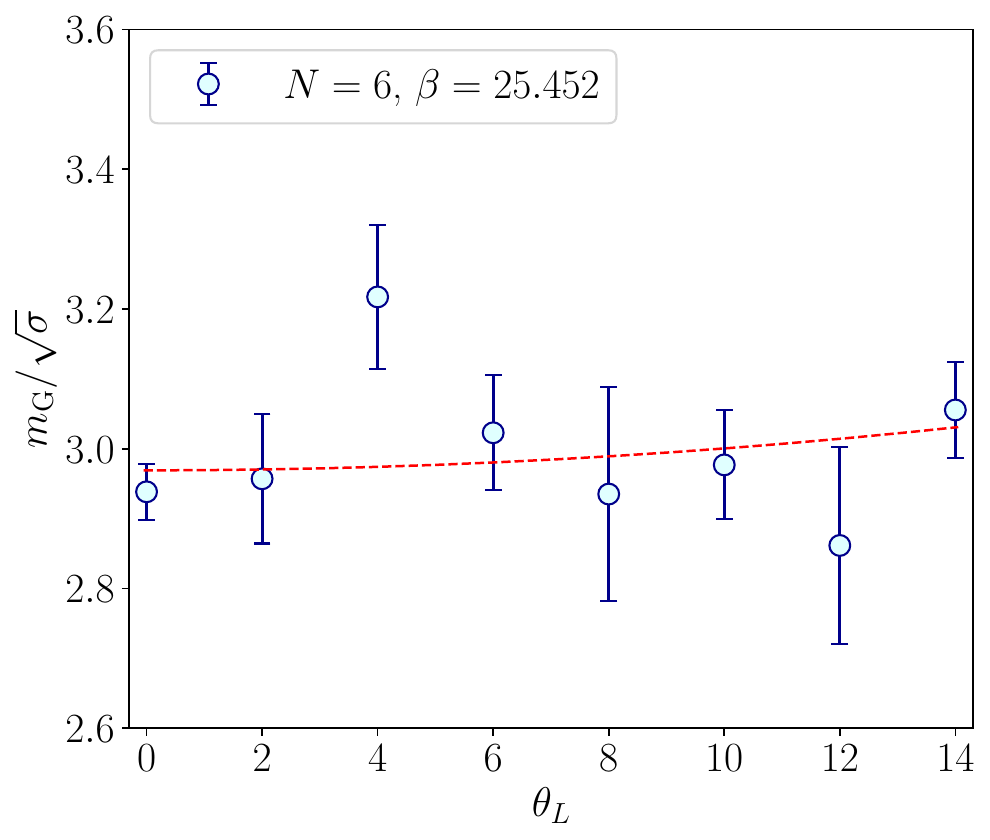}
\caption{Results for $N=6$ with $\beta=25.452$. Examples of fits performed by
using the functional form Eq.~\eqref{eq:fit_function} for several quantities:
(top left panel) the lightest glueball mass $a m_{\G}$, (top right panel) the
string tension $a^2 \sigma$, (bottom panel) the ratio
$m_{\G}/\sqrt{\sigma}$.}
\label{fig:ex_fit_theta_N6}
\end{figure}

Our results at $N=6$ for $m_2$ and $s_2$ are summarized in
Tab.~\ref{tab:data_N6}, while all raw data for $m_{\G}(\theta_L)$ and
$\sigma(\theta_L)$ can be found in App.~\ref{sec:app_raw_data}. As it is clear
from data in Tab.~\ref{tab:data_N6}, and from the example shown in
Fig.~\ref{fig:ex_fit_theta_N6}, the $\theta$ dependence of $m_\G(\theta)$ and
$\sigma(\theta)$ is much milder at $N=6$ than it is at $N=3$. In particular, 
we can only provide upper bounds for $m_2$. This behavior is compatible with the
expectation that $m_2$ and $s_2$ are suppressed in the large-$N$ limit. 

\begin{figure}[!t]
\centering
\includegraphics[scale=0.435]{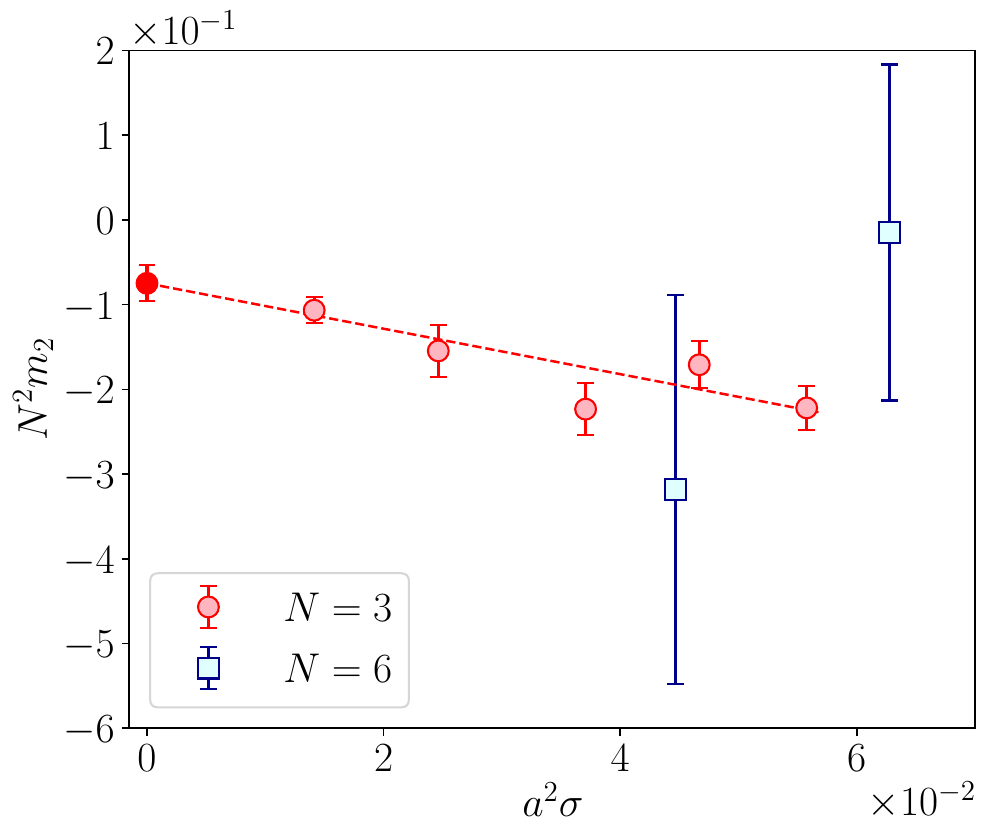}
\includegraphics[scale=0.435]{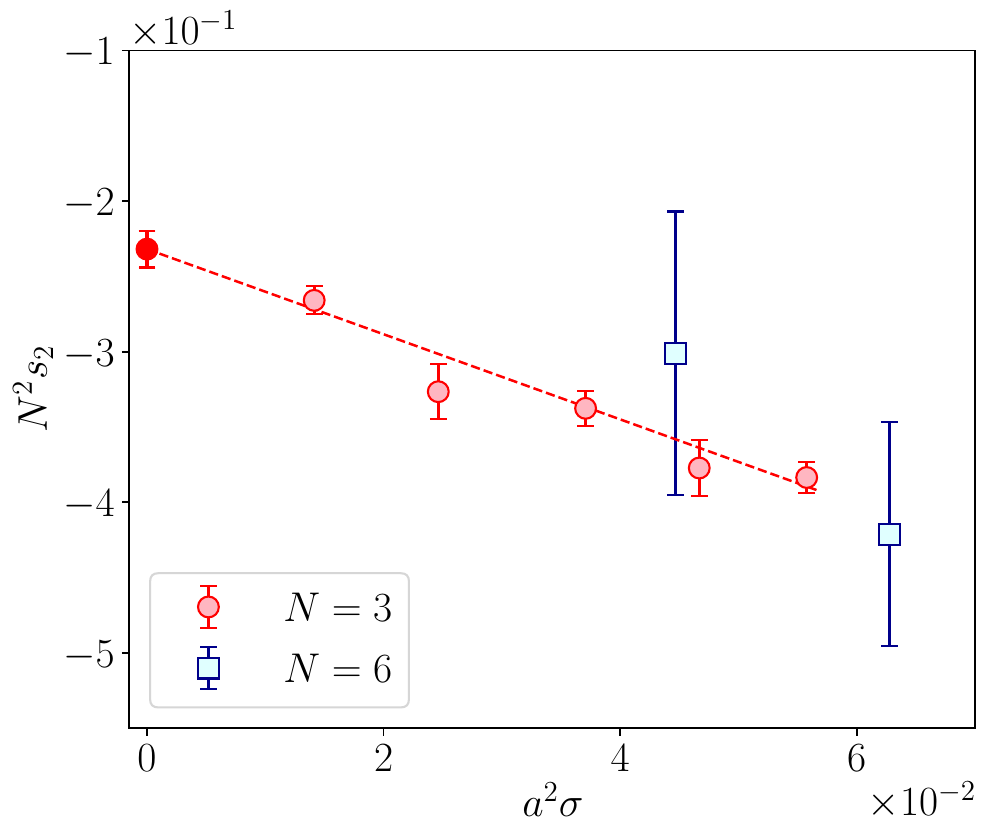}
\caption{Continuum scaling of $N^2 m_2$ and $N^2 s_2$ for $N=3$ and $N=6$.} 
\label{fig:ex_cont_limit_N6}
\end{figure}

The accuracy of the data collected at $N=6$ is not sufficient to 
directly test the expected~\cite{DelDebbio:2006yuf} $1/N^2$ behavior of 
$m_2$ and $s_2$. However, our data are definitely consistent with this scaling law, 
as can be appreciated from Fig.~\ref{fig:ex_cont_limit_N6}, where 
$N^2 m_2$ and $N^2 s_2$
are plotted together for $N=3$, $6$ as a function of $\sigma a^2$.  
To describe the large-$N$ behavior we can define the parameters $\bar{s}_2$ and $\bar{m}_2$ by:
\begin{equation} \label{eq:bar_def}
s_2 \simeq \bar{s}_2 / N^2+\mathcal{O}(N^{-4})\ ,\quad
m_2 \simeq \bar{m}_2 /N^2+\mathcal{O}(N^{-4})~.
\end{equation}
Assuming the leading order large-$N$ scaling to be accurate already for $N\ge 3$,
as is the case for other $\mathcal{O}(\theta^2)$ quantities~\cite{Bonanno:2020hht,Bonanno:2023hhp}, 
we can estimate these parameters using the results obtained for $N=3$:
\begin{equation}
\bar{s}_2 \simeq -0.23(1)\ ,\quad
\bar{m}_2 \simeq -0.075(20)\ .
\end{equation}

\begin{table}[!t]
\begin{center}
\begin{tabular}{|c|c|l|c|c|l|c|}
\hline
$L$ & $\beta$ & \multicolumn{1}{c|}{$Z_Q$} & $a m_{\zpp}$ & $m_2$ & \multicolumn{1}{c|}{$a^2 \sigma$} & $s_2$ \\

\hline
14 & 25.056 & 0.12053(88)* & 0.7369(75) & -0.0004(55) &  0.06278(27) & -0.0117(21) \\
\hline
16 & 25.452 & 0.13834(46)  & 0.6297(64) & -0.0088(64)   &  0.04467(25) & -0.0084(26) \\
\hline
\end{tabular}
\end{center}
\caption{Summary of the results obtained for $N=6$. The value of $Z_Q$ denoted
by an asterisk is from Ref.~\cite{Bonanno:2020hht}, while the value for
$\beta=25.452$ has been computed anew in this work.}
\label{tab:data_N6}
\end{table}

\section{Conclusions}\label{sec:conclu}

In this paper we presented a novel investigation of the $\theta$-dependence of
the spectrum of four-dimensional $\SU(N)$ Yang--Mills theories. In particular,
we focused on the leading order $\mathcal{O}(\theta^2)$ dependence of the string tension
$\sigma(\theta)$ and of the lightest glueball state $m_\G(\theta)$, 
and estimated the value of the parameters $s_2$ and $m_2$ 
defined in Eq.~\eqref{eq:s2m2_def}. The present
study has been carried out by means of lattice simulations performed at
imaginary values of the topological $\theta$-angle, using the so called
\emph{analytic continuation} approach. This approach proved to be 
extremely effective in reducing statistical uncertainties with respect to 
the previously adopted Taylor expansion method.

Our main results have been obtained for the $N=3$ theory, for which we were able to perform a
controlled continuum extrapolation of $m_2$ and $s_2$, with the results:
\beq
m_2(N=3) &=& -0.0083(23),\\
s_2(N=3) &=& -0.0258(14).
\eeq 

The case $N=6$ was much more challenging, as expected \emph{a priori}. Larger
auto-correlation times make numerical simulations particularly demanding.
Moreover, the values of $m_2$ and $s_2$ at $N=6$ are expected (by large-$N$
scaling) to be suppressed by a factor $\approx 4$ with respect to their value
at $N=3$. To address the first problem we used the 
Parallel Tempering on Boundary Conditions algorithm, and yet
the best we could do was to estimate the values of $m_2$ and
$s_2$ for only two values of the lattice spacing, although two
quite fine ones. 

Due to the limited accuracy of the results at $N=6$, we cannot perform a
stringent test of the expected $\mathcal{O}(N^{-2})$ behavior of $m_2$ and $s_2$.
However, our data are definitely consistent with this expectation. Based on our previous
experiences with $\mathcal{O}(\theta^2)$ corrections (see, e.g.,
Refs.~\cite{Bonanno:2020hht,Bonanno:2023hhp}), it is natural to expect the
large-$N$ scaling of $m_2$ and $s_2$ to be well approximated by the leading
$\mathcal{O}(N^{-2})$ behavior already for $N=3$; using this assumption we get for the
coefficients $\bar{s}_2$ and $\bar{m}_2$ defined in Eq.~\eqref{eq:bar_def} the
estimates
\begin{equation}\label{eq:bar_res}
\bar{s}_2 \simeq -0.23(1)\ ,\quad
\bar{m}_2 \simeq -0.075(20)\ .
\end{equation}

As noted in the introduction, the parameters $m_2$ and $s_2$, which parameterize
the leading order $\theta$-dependence of the spectrum, also parameterize the way
in which the lightest glueball mass and the string tension are affected by the
topological freezing, i.e., the systematics induced by using for their
estimation an ensemble of gauge configurations with fixed topological charge.
Such a quantitative information is very useful. Indeed, given the very fast
growth of the integrated auto-correlation time of the topological charge as the
continuum limit is approached, it is quite common to 
perform simulations at fixed $Q=0$ with large-$N$ gauge groups.
Using our results in Eq.~\eqref{eq:bar_res} and the general formula
Eq.~\eqref{eq:Qbias}, it is possible to estimate the bias induced by using a
fixed topological background. It turns out that already modestly large volumes 
are sufficient to have a negligible bias: considering for example the case $N=3$, 
in which case $\chi^{1/4}$ is roughly equal to $1\text{ fm}^{-1}$ (see, e.g.,~\cite{Bonanno:2023ple}), we have
\beq
\frac{\Delta m_{\zpp}}{m_{\zpp}}\Bigg\vert_{N=3} \approx \frac{m_2}{2\chi V} \approx -0.08\%
\eeq
for $V\simeq(1.5\text{ fm})^4$. Using larger values of $N$ this estimate
becomes drastically more favorable, since the topological susceptibility
changes only slightly (see, e.g., Refs.~\cite{Bonati:2016tvi,Ce:2016awn,Bonanno:2020hht}), while $m_2$ scales as $1/N^2$. These estimates constitute an independent
confirmation of the results obtained in Ref.~\cite{Bonanno:2022yjr}, with the
advantage of providing a quantitative upper bound to the accuracy that can be
achieved when using simulations at fixed topological sector.

The results presented in this paper can be extended quite naturally in several
different ways. One possibility is to accurately investigate $s_2$ and $m_2$ for
$N>3$, in order to quantitatively asses the $N$ dependence of these
coefficients.  From the previous discussion it should be clear that this is not
an easy task, and some ideas are required to further improve the signal-to-noise 
ratio. A second possibility is to study the excited glueball spectrum. In
particular, it would be very interesting to understand (both theoretically and
numerically) how the $\mathcal{O}(\theta^2)$ correction to the mass depends on 
the state considered. Indeed, these corrections can not be all independent from each
other, since they are indirectly related to the $\theta$-dependence of the free
energy in the confined phase. This can be easily understood by using hadron
resonance gas models like, e.g., those discussed in~\cite{Caselle:2015tza,
Trotti:2022knd}, where it was shown that a determination of the glueball masses
is sufficient to obtain quantitatively accurate estimates of thermodynamical
quantities. Finally, it would be very interesting to study models in which
$s_2$ and $m_2$ or, more generally, the $\theta$-dependence of the spectrum, can
be investigated analytically (and is non-trivial). Two-dimensional 
$\mathrm{CP}^{N-1}$ models are natural candidates.

\section*{Acknowledgements}

The work of C.~Bonanno is supported by the Spanish Research Agency (Agencia
Estatal de Investigación) through the grant IFT Centro de Excelencia Severe
Ochoa CEX2020- 001007-S and, partially, by grant PID2021-127526NB-I00, both
funded by MCIN/AEI/ 10.13039/ 501100011033. C.~Bonanno also acknowledges
support from the project H2020-MSCAITN-2018-813942 (EuroPLEx) and the EU
Horizon 2020 research and innovation programme, STRONG-2020 project, under
grant agreement No 824093. 
The work of D.~Vadacchino is supported by STFC under Consolidated Grant No.~ST/X000680/1.
Numerical calculations have been performed on the
\texttt{Galileo100} machine at Cineca, based on the project IscrB\_ITDGBM, on
the \texttt{Marconi} machine at Cineca based on the agreement between INFN and
Cineca (under project INF22\_npqcd), and on the Plymouth University cluster.

\appendix

\section*{Appendix}

\section{Two-dimensional \texorpdfstring{$\mathrm{U}(N)$}{U(N)} Yang--Mills theories}\label{sec:UN2d}

Two-dimensional Yang--Mills theories are particularly simple to investigate in the
thermodynamic limit: neglecting boundary conditions, we can fix $U_1(x)=1$ 
on all the sites and $U_2(x)=1$ along a single line at constant $x_1$. 
In this way it is simple to show, using the invariance
properties of the Haar measure, that link integrals can be traded for plaquette
integrals and the theory reduces to a single-plaquette model~\cite{Gross:1980he}.

Using for the topological charge density the definition~\cite{Bonati:2019ylr}
\begin{equation}
q(x)=\frac{1}{2\pi}\mathrm{arg}\det(\Pi_{12}(x))\ ,
\end{equation}
where $\Pi_{12}(x)$ denotes the plaquette in position $x$, the string tension at inverse 't Hooft lattice coupling $\lambda=\beta/(2N^2)$ can be written as~\cite{Gross:1980he}
\begin{equation}\label{eq:sigmaN}
\begin{aligned}
\sigma(N, \lambda, \theta)&=-\log\left(\frac{1}{Z_{1p}(N,\lambda,\theta)}
\int \mathrm{d}W\frac{1}{N}\mathrm{tr}(W)e^{N\lambda\mathrm{tr}(W+W^{\dag})+\frac{\theta}{2\pi}\mathrm{tr}\log(W)}\right)\\
&=-\log\left(\frac{1}{2N^2}\frac{\mathrm{d}}{\mathrm{d}\lambda}\log Z_{1p}(N,\lambda,\theta)\right)\ ,
\end{aligned}
\end{equation}
where $\mathrm{d}W$ is the Haar measure on U($N$) and
\begin{equation}
Z_{1p}(N,\lambda,\theta)=\int \mathrm{d}We^{N\lambda\mathrm{tr}(W+W^{\dag})+\frac{\theta}{2\pi}\mathrm{tr}\log(W)}\ . 
\end{equation}
It is also simple to show that the connected correlator of two plaquette identically 
vanishes whenever the two plaquettes are not coincident, hence no finite glueball mass 
can be defined.

Using the Weyl form of the Haar measure for class functions, it is possible to rewrite, 
using manipulations completely analogous to those used in~\cite{Gross:1980he, Bars:1979xb}, 
the single-plaquette partition function as a $N\times N$ determinant~\cite{Bonati:2019ylr}
\begin{equation}\label{eq:Zdet}
Z_{1p}(N,\lambda,\theta)=\det\left(\mathcal{I}_{i-j+\frac{\theta}{2\pi}}(2N\lambda)\right)\ ,\quad i,j=1,\ldots N\ ,
\end{equation}
where the functions $\mathcal{I}_{\nu}(x)$ are defined by
\begin{equation}
\mathcal{I}_{\nu}(x)=\frac{1}{2\pi}\int_{-\pi}^{\pi} e^{i\nu\phi} e^{x\cos\phi}\mathrm{d}\phi\ .
\end{equation}
We thus have (for $k,j=1,\ldots,N$)
\begin{equation}
Z_{1p}(N,\lambda,\theta)=\int\mathrm{det}\left[e^{i\phi_j(k-j)}\right]e^{i\frac{\theta}{2\pi}\sum_j\phi_j} e^{2N\lambda\sum_j\cos\phi_j}\prod\frac{\mathrm{d}\phi_j}{2\pi}\ ,
\end{equation}
and to study the leading behavior in the limit $\lambda\to\infty$ it is sufficient to replace $\cos\phi_j$ by $1-\frac{1}{2}\phi_j^2$ in the exponentials, obtaining
\begin{equation}
Z_{1p}(N,\lambda\gg 1,\theta)=
\left(\frac{e^{2\lambda N}}{\sqrt{4\pi\lambda N}}\right)^N 
\det\left[e^{-\frac{1}{4N\lambda}\left(k-j+\frac{\theta}{2\pi}\right)^2}\right]\ .
\end{equation}
By using the multi-linearity of the determinant we can rewrite this expression as follows
\begin{equation}
\begin{aligned}
Z_{1p}(N,\lambda\gg 1,\theta)&= \left(\frac{e^{2\lambda N}}{\sqrt{4\pi\lambda N}}\right)^N 
\det\left[e^{-\frac{1}{4N\lambda}\left(k-j\right)^2}\right]\exp\left\{-\frac{1}{4\lambda}\left(\frac{\theta}{2\pi}\right)^2\right\}=\\
&=Z_{1p}(N,\lambda\gg 1,\theta=0)\exp\left\{-\frac{1}{4\lambda}\left(\frac{\theta}{2\pi}\right)^2\right\}\ ,
\end{aligned}
\end{equation}
where
\begin{equation}
Z_{1p}(N,\lambda\gg 1,\theta=0)=\left(\frac{e^{2\lambda N}}{\sqrt{4\pi\lambda N}}\right)^N e^{-\frac{1}{2\lambda N}\sum_{k=1}^N k^2}
\det\left[e^{-\frac{jk}{2N\lambda}}\right]\ ,
\end{equation}
and the remaining determinant can be related to a Vandermonde determinant.
Using these expressions in Eq.~\eqref{eq:sigmaN} we see that for $\lambda\gg 1$ we have
\begin{equation}
\frac{\mathrm{d}}{\mathrm{d}\lambda}\log Z_{1p}(N,\lambda\gg 1,\theta=0)= 2N^2-\frac{N}{2\lambda}+o(\lambda^{-1})\ ,
\end{equation}
and to reliably estimate subleading terms we should go beyond the leading order expansion 
of Eq.~\eqref{eq:Zdet}. The $\theta\neq 0$ contribution is the subleading $N$-independent correction $\frac{\theta^2}{16\pi^2\lambda^2}$, hence the continuum string tension does 
not depend on $\theta$.

\section{Raw data for four-dimensional \texorpdfstring{$\SU(3)$}{SU(3)} and \texorpdfstring{$\SU(6)$}{SU(6)} Yang--Mills theories}\label{sec:app_raw_data}

In this appendix we collect all the results obtained for $am_\G(\theta_L)$ and 
$a^2\sigma(\theta_L)$ at the different values of the inverse lattice coupling $\beta$
for $N=3$ (Tab.~\ref{tab:raw_data_N3}) and $N=6$ (Tab.~\ref{tab:raw_data_N6}).

\begin{table}[!t]
\small
\begin{center}
\begin{tabular}{|c|c|c|}
\hline
$\theta_L$ & $a m_{\G}$ & $a^2 \sigma$ \\
\hline
\hline
\multicolumn{3}{|c|}{$N=3$, $L=16$, $\beta=5.95$}\\
\hline
0  & 0.7487(62) & 0.05576(19) \\
2  & 0.7409(15) & 0.05478(43) \\
4  & 0.7560(15) & 0.05652(23) \\
6  & 0.7473(14) & 0.05718(20) \\
8  & 0.7589(72) & 0.05816(45) \\
10 & 0.804(16)  & 0.05961(23) \\
12 & 0.7892(71) & 0.06097(22) \\
14 & 0.7922(77) & 0.06283(24) \\
16 & 0.8231(76) & 0.06515(26) \\ 
\hline
\hline
\multicolumn{3}{|c|}{$N=3$, $L=18$, $\beta=6.00$}\\
\hline
0  & 0.6865(61) & 0.04659(31) \\
5  & 0.7008(65) & 0.04757(17) \\
8  & 0.744(12)  & 0.04930(19) \\
10 & 0.7190(67) & 0.05014(19) \\
12 & 0.7301(62) & 0.05155(21) \\
14 & 0.7364(64) & 0.05475(46) \\
\hline
\hline
\multicolumn{3}{|c|}{$N=3$, $L=18$, $\beta=6.07$}\\
\hline
0  & 0.6261(45) & 0.03710(11) \\
5  & 0.6296(55) & 0.03787(14) \\
8  & 0.6417(55) & 0.03898(24) \\
10 & 0.6495(57) & 0.04027(14) \\
12 & 0.676(11)  & 0.04149(15) \\
14 & 0.707(11)  & 0.04357(28) \\
\hline
\hline
\multicolumn{3}{|c|}{$N=3$, $L=22$, $\beta=6.20$}\\
\hline
0  & 0.5187(78)  & 0.02425(23) \\
5  & 0.5312(78)  & 0.02543(15) \\
8  & 0.5410(85)  & 0.02653(10) \\
10 & 0.5482(83)  & 0.02760(29) \\
12 & 0.5576(92)  & 0.02839(30) \\
14 & 0.5792(98)  & 0.03036(20) \\
\hline
\hline
\multicolumn{3}{|c|}{$N=3$, $L=30$, $\beta=6.40$}\\
\hline
0  & 0.3923(36) & 0.014104(80) \\
2  & 0.4014(40) & 0.014297(87) \\
4  & 0.4033(44) & 0.014304(93) \\
6  & 0.4033(46) & 0.014837(97) \\
8  & 0.4084(48) & 0.015314(97) \\
10 & 0.4154(46) & 0.01622(16)  \\
12 & 0.4254(47) & 0.016680(83) \\
14 & 0.4397(79) & 0.01770(13)  \\
\hline
\end{tabular}
\end{center}
\caption{Summary of all obtained results for $N=3$.}
\label{tab:raw_data_N3}
\end{table}

\clearpage

\begin{table}[!t]
\small
\begin{center}
\begin{tabular}{|c|c|c|}
\hline
$\theta_L$ & $a m_{\G}$ & $a^2 \sigma$ \\
\hline
\hline
\multicolumn{3}{|c|}{$N=6$, $L=14$, $\beta=25.056$}\\
\hline
0  & 0.748(11) & 0.06137(77) \\
2  & 0.757(27) & 0.06351(44) \\
4  & 0.761(27) & 0.06307(44) \\
6  & 0.717(15) & 0.06261(97) \\
8  & 0.725(13) & 0.06241(99) \\
10 & 0.679(25) & 0.06256(95) \\
12 & 0.811(31) & 0.06426(50) \\
14 & 0.742(13) & 0.06559(47) \\
16 & 0.735(16) & 0.06530(45) \\
\hline
\hline
\multicolumn{3}{|c|}{$N=6$, $L=16$, $\beta=25.452$}\\
\hline
0  & 0.6246(78) & 0.04518(48) \\
2  & 0.624(19)  & 0.04450(43) \\
4  & 0.676(19)  & 0.0441(13)  \\
6  & 0.643(17)  & 0.04519(46) \\
8  & 0.622(32)  & 0.04496(38) \\
10 & 0.632(16)  & 0.04507(48) \\
12 & 0.610(30)  & 0.04544(46) \\
14 & 0.658(15)  & 0.04633(37) \\
\hline
\end{tabular}
\end{center}
\caption{Summary of all obtained results for $N=6$.}
\label{tab:raw_data_N6}
\end{table}

\providecommand{\href}[2]{#2}\begingroup\raggedright\endgroup

\end{document}